\documentclass[12pt,reqno]{amsart}
\usepackage{amsfonts}
\usepackage{amssymb}
\usepackage[english]{babel}
\usepackage{mathrsfs}

\renewcommand\a{\alpha}
\newcommand\af{\frak{a}}
\newcommand{\g}{{\mathfrak g}}
\newcommand\op[1]{\mathop{\rm #1}\nolimits}
\newcommand\Qq{{\let\mathcal\mathscr\mathcal Q}}
\newcommand{\R}{{\mathbb R}}
\newcommand\s{\mathfrak{s}}
\newcommand\si{\mathfrak{s}_\infty}

\begin{document}

\title[B\"acklund Transformation between two 4D equations]{%
A B\"acklund transformation between\\
4D Mart{\'{\i}}nez Alonso -- Shabat
and Ferapontov -- Khusnutdinova  equations}

\author[Boris Kruglikov \& Oleg Morozov]{Boris Kruglikov $^\dag$ and Oleg Morozov $^\ddag$}

\address{$^\dag$~
Institute of Mathematics and Statistics, NT-Faculty, University of Troms\o,
Troms\o\ 90-37,  Norway. E-mail: boris.kruglikov@uit.no}

\address{$^\ddag$~
Faculty of Applied Mathematics, AGH University of Science and Technology, Al. Mickiewicza 30,
Krak\'ow 30-059, Poland. E-mail: morozov{\symbol{64}}agh.edu.pl}

\maketitle

The aim of this note is to construct a B\"acklund transformation between the Lax-integrable 4-dimensional equations
 \begin{equation}
u_{ty} = u_z\,u_{xy} - u_y\,u_{xz}
\label{MASh4}
 \end{equation}
and
 \begin{equation}
u_{yz} = u_{tx} + u_x\,u_{xy} - u_y\,u_{xx}
\label{FKh4}
 \end{equation}
introduced in \cite{MASh} and \cite{FKh}, respectively.
Equation (\ref{FKh4}) has the following Lax pair \cite{FKh}:
 \begin{equation}
\left\{
\begin{array}{lcl}
v_t &=& \lambda \, v_y + u_y \, v_x,\\
v_z &=& (u_x+\lambda)\,v_x.
\end{array}
\right.
\label{FKh4_covering}
 \end{equation}
with non-removable parameter $\lambda$. Excluding $u$ from (\ref{FKh4_covering})
and normalizing $\lambda=1$ in the resulting equation by re-scaling we get the equation
 \begin{equation}
v_x\,v_{yz} = v_x\,v_{tx}+(v_y - v_t)\,v_{xx} - (v_x-v_z)\,v_{xy}.
 \label{mFKh4}
\end{equation}
Thus the differential covering (\ref{FKh4_covering}) (in terms of  \cite{KV84}) defines
a B\"acklund transformation between equations (\ref{FKh4}) and (\ref{mFKh4}).

\vskip 7 pt
\noindent
{\sc Proposition}. {\it Equations (\ref{MASh4}) and (\ref{mFKh4}) are point equivalent.}

\vskip 7 pt
\noindent
{\it Proof}.
Prolongation of the transformation $\psi \colon \R^4\times\R \to \R^4\times\R$
 \begin{equation}
\psi(t,x,y,z,v)=(t,z,u,y+t,-x)
\label{BT}
 \end{equation}
maps equation (\ref{mFKh4}) to equation (\ref{MASh4}).
\hfill {\sc q.e.d.}

\vskip 10 pt
\noindent
{\sc Corollary}. {\it The superposition of (\ref{BT}) and (\ref{FKh4_covering}) defines a
B\"acklund transformation be\-twe\-en equations (\ref{MASh4}) and (\ref{FKh4}).}

\medskip

Let us explain how we came to the existence of this transformation. We computed the
contact symmetries of some integrable equations.

The infinite part of the symmetry group is parametrized by 3 copies of group
$\op{Diffeo}(\R^1)\,\hat\otimes \, C^\infty(\R^1)$ consisting of 1-parametric dif\-fe\-o\-mor\-phisms of $\R^1$
(can be changed to $S^1$); its Lie algebra is identified with the space of 1-parametric vector
fields in 1D: $\Qq=\{a(x,y)\partial_x\}\simeq C^\infty(\R^2)$ (here $\R^2=T\R^1$ can be changed to $TS^1=S^1\times\R^1$)
with the Lie bracket
 $$
[a(x,y),b(x,y)]=a\, b_x-a_xb.
 $$
Everywhere below $\af_i\simeq\Qq$ will be the graded pieces of the infinite part $\si$ of the symmetry algebra $\g$.
The upper index indicates the different commuting copies of the graded subalgebras, i.e. $[\g_i^\a,\g_j^\beta]\subset\delta^{\a\beta}\g_{i+j}^\a$.

The symmetry algebra of equation (\ref{MASh4}) is the semi-direct product
$\g=\s_\diamond\ltimes\si$, where ($\af_i\simeq\Qq$)
 $$
\si=(\af'_0\oplus\af_1')\oplus\af_0'';\qquad \s_\diamond=(\R^1\ltimes\R^1)\oplus(\R^1\ltimes\R^1).
 $$

Similarly, the symmetry algebra of equation (\ref{FKh4}) is the semi-direct product
$\g=\s_\diamond\ltimes\si$, where
 $$
\si=\af_0\oplus\af_1\oplus\af_2;\qquad \s_\diamond=\mathfrak{sl}_2\ltimes(\R^2\rtimes\R^1).
 $$

The modified (B\"acklund equivalent) version of equation (\ref{MASh4}) is the following
equation, \cite{MorozovSergyeyev2014}, (we again normalize $\lambda=1$ by re-scaling)
 \begin{equation}
u_xu_{ty}=(u_t+u_z)\,u_{xy}-u_y\,u_{xz}.
\label{mMASh4}
 \end{equation}
Its symmetry algebra is the semi-direct product $\g=\s_\diamond\ltimes\si$, where
 $$
\si=\af_0'\oplus\af_0''\oplus\af_0''';\qquad \s_\diamond=(\R^2\rtimes\R^1);
 $$
The equations (\ref{MASh4}), (\ref{FKh4}), (\ref{mMASh4}) have different symmetry algebras.
But equation (\ref{mFKh4}), which is a modified version of (\ref{FKh4}), has the same symmetry as (\ref{MASh4}).
Although coincidence of symmetry algebras is only a necessary condition for equivalence of two equations,
in this case we obtain the equivalence defined by (\ref{BT}).

\bigskip
\noindent
{\sc Remark}. In \cite{KM} we constructed symmetric integrable deformations of some heavenly type
equations. These also exist for the equations (\ref{MASh4}), (\ref{FKh4}), (\ref{mFKh4}) and (\ref{mMASh4})
considered in this paper. For the first of them we have such an integrable deformation:
 $$
u_{ty}-u_zu_{xy}+u_yu_{xz}+(Q\,u_y)_y=0,\quad Q=Q(t,y,z).
 $$
This equation has a covering defined by system
 \[
\left\{
\begin{array}{lcl}
w_y &=& \lambda \,u_y \, w_x,
\\
w_z &=& \lambda \, \left((u_z-\lambda\,Q\,u_y)\,w_x - w_t\right).
\end{array}
\right.
 \]
The functional parameter $Q$ is (essentially) non-removable.
Details of this construction and information on other integrable deformations will appear in a
forthcoming paper.


\begin{thebibliography}{99}

\bibitem{FKh} E.\,V.\ Ferapontov, K.\,R.\ Khusnutdinova, {\it Hydrodynamic reductions of multi-dimensional
        dispersionless PDEs: the test for integrability\/}, Journal of Mathematical Physics 45 (6), 2365-2377 (2004).

\bibitem{KV84} I.\,S.\ Krasil'shchik, A.\,M.\ Vinogradov, {\it Nonlocal symmetries and the theory of
        coverings\/}, Acta Appl. Math. {\bf 2}, 79--86 (1984).

\bibitem{KM} B.\ Kruglikov, O.\ Morozov, {\it Integrable dispersionless PDE in 4D, their symmetry pseudogroups
        and deformations\/}, arXiv:1410.7104 (2014).

\bibitem{MASh}
    L.\ Mart{\'{\i}}nez Alonso, A.\,B.\ Shabat, {\it Hydrodynamic reductions and solutions of a universal
    hierarchy\/}, Theoret. Math. Phys. {\bf 104}, 1073--1085 (2004).

\bibitem{MorozovSergyeyev2014}
    O.I. Morozov, A. Sergyeyev, {\it  The four-dimensional Mart{\'{\i}}nez Alonso--Shabat equation:
    Reductions and nonlocal symmetries}, Journal of Geometry and Physics {\bf 85} 40--45 (2014)


\end{thebibliography}
\end{document}